\documentclass[nocite]{epl}
\usepackage{graphicx}

\title{Deformation of grain boundaries in polar ice}
\shorttitle{Deformation of grain boundaries in polar ice}

\author{
G. Durand\inst{1}, F. Graner\inst{2}\protect\cite{cauthor}, and
J. Weiss\inst{1}} \shortauthor{G. Durand \etal}
\institute{\inst{1} Laboratoire de Glaciologie et de G\'eophysique
de l'Environnement (FRE  2192 CNRS - Universit\'{e} J. Fourier
Grenoble 1), BP  96, F-38402 Saint Martin d'H\`eres cedex, France\\
         \inst{2}
        Laboratoire de Spectrom\'etrie Physique (UMR 5588
CNRS -Universit\'{e} J. Fourier Grenoble 1), BP 87, F-38402 Saint Martin
d'H\`eres cedex,  France
}

\begin{document}

\maketitle

\date{\today}

\begin{abstract}

The ice microstructure (grain boundaries) is a key feature used to
study ice evolution and to investigate past climatic changes. We
studied a deep ice core, in Dome Concordia, Antarctica, which records
past mechanical deformations. We measured
a ``texture tensor'' which characterizes the pattern geometry and
reveals local heterogeneities of deformation along the core.
These
results question key assumptions of the current models used for
dating.
\end{abstract}

\bigskip

\section{Motivations}

Polar ice cores are the focus of many investigations because they
record the history of climatic changes. Owing to snow
accumulation, snow to ice transformation and slow ice sheet flow
($\sim$ 10$^{-12} { } s^{-1}$), a journey down to the deep layers of
the ice sheet is a journey back to several hundred of thousands of
years into the past \cite{vostok}.

A crucial step of paleoclimatic studies from ice cores is dating.
In Antarctica, counting annual layers is impossible
\cite{schwander}: absolute dating is only possible for the very
top of the ice cores where ice layers containing volcanic
impurities can be related to historical volcanic eruptions. Below,
dating relies on ice sheet flow models of the
evolution of ice layer thinning with depth \cite{schwander}. Such
models are loosely constrained by the identification of large
climatic transitions. For the sake of
simplicity, these models assume a smooth and monotonous increase
of the thinning with depth, hence ignore any possible localization
of the deformation \cite{schwander,hammer}.

In this letter, we question this essential assumption.
We present a method to extract geometrical
information (such as thinning, shear, localization of the
deformation) from pictures of a cellular pattern using local spatial
averages of the ``texture tensor"
\cite{aubouy}. We apply this analyzis
to the grain boundaries
(the
so-called ``microstructure") of ice samples from a deep
ice core.

\section{Samples}
Dome Concordia, Antarctica (75$^\circ$ 06' 04" S, 123$^\circ$ 20'
52" E, elevation 3233 m a.s.l) is
       at the summit of an Antarctic ice
dome. It has been chosen  because it
is usually assumed that the ice flow is axisymmetric around the
vertical ($z$) axis, and isotropic within the horizontal
plane.
In February 2003, the
European ice core drilling program (EPICA) reached the depth of
3201.65 m below the ice surface, close to the ground (depth 3309
${\pm}$  22 m).

\begin{figure}
\begin{center}
\onefigure[width=10cm]{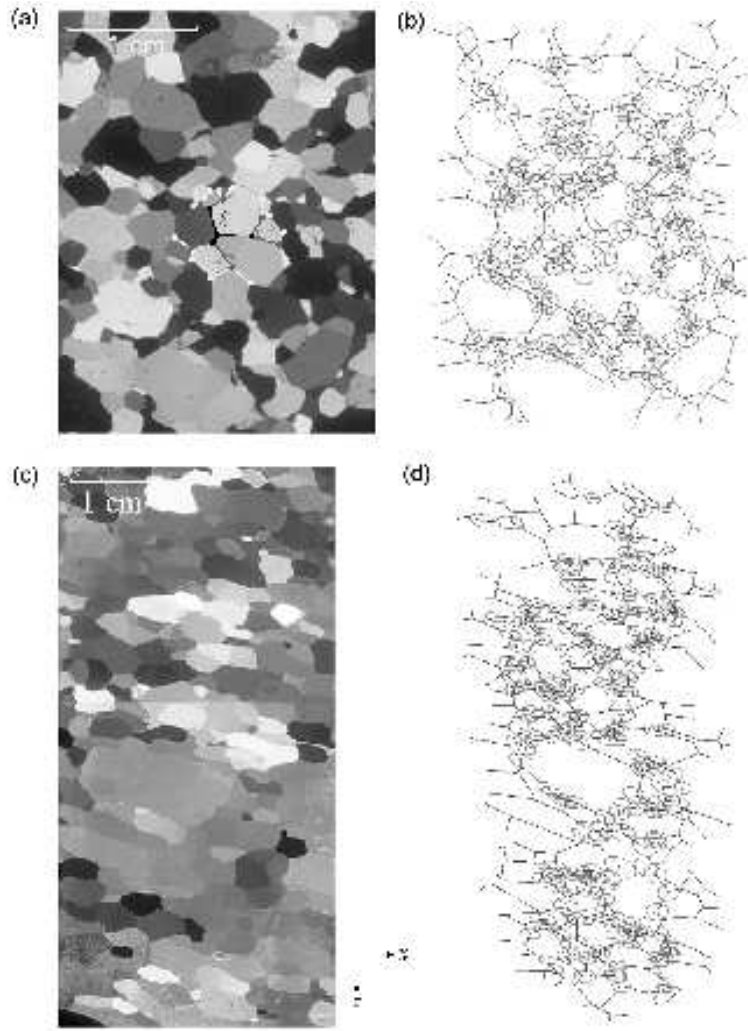}
\end{center}
\caption{ Two thin sections of ice imaged between crossed
polarizers in white light (a and c) and their corresponding
$\overline{\overline{M}}$ tensor analysis (b and d).
Each grain has an almost uniform crystallographic
orientation, visualized by its color. (a) At 362 m depth, the
microstructure looks isotropic; (c) at 2629 m depth the
microstructure is visibly anisotropic.
Superimposed on (a): notations used in the measurement of
$\overline{\overline{M}}$, here {\it e.g.} at the site labeled by a black dot.
       We note $\vec \ell$ a vector linking it to one of its neighbours.
There are 3 such vectors in the first shell, $p=1$ (black); 6 in
shell $p=2$ (grey); 12 in shell $p=3$ (white).
The texture tensor
$\overline{\overline{M}} $ measured at each site is represented as
an ellipse (362 m: (b) and 2629 m: (d)), with its axes along the 
eigenvectors, and its
half-axes proportional to eigenvalues $\lambda_1 , \lambda_2$
respectively. A site around which the pattern is isotropic is
represented by a circle; conversely, a strong anisotropy is
represented as an elongated ellipse. The size of the ellipse (same
scale for each ellipse) represents the local length of the grain
boundaries, {\it i.e} quantifies the local grain size. Due to the
definition of $\overline{\overline{M}} $ (eq. \protect\ref{defM}), we
exclude sites closer than $p$ grains
from the image boundary; here $p=3$.}
\label{polariseurs}
\end{figure}

Samples have been extracted and
transported according to standard procedures \cite{gay}. We
analyzed $\sim$11~cm high, $\sim 0.1$~mm thin sections (Fig.
\ref{polariseurs}), with a sampling
interval varying from 2 to 25 m.  The vertical axis on a thin
section always corresponds to the true ice core vertical axis
$z$; its horizontal axis, labelled $x$ in what follows,
has an unknown and variable orientation within the horizontal plane.
The tilt of the core axis from the {\itshape in situ} vertical axis 
was below 0.5$^\circ$ above
$z= 2300$ m, then increased up to 5.2$^\circ$ at 3100 m.

\section{Data analyzis}

By processing  pictures
of ice under crossed
polarizers, we
determine the grain boundaries \cite{gay}, then the sites (``vertices")
where three boundaries meet, see Fig. (\ref{polariseurs}a). For each
pair of neighbouring sites, we draw the vector which links them and denote
it by $\vec{\ell}$. We then construct the tensor $\vec{\ell}\otimes
\vec{\ell}$: its coordinates are $(\ell _{i}\ell _{j})$, where
$i,j$ are here $x$ or $z$; this tensor is not sensitive to the
sign of the vector (it is invariant under $\vec{\ell} \to -
\vec{\ell}$), but it characterizes its length $\ell$ and its
direction.


To   statistically  characterize the
      pattern,
ref. \cite{aubouy} proposed to  define the {\it texture tensor}
$\overline{\overline{M}}$,  as the
average of
$\vec{\ell}\otimes \vec{\ell}$ over a box of fixed
size.  Such box should be smaller than the image (in order to
visualize local
details), but still
large enough to include a number $N\gg 1$ of wall vectors (to
have relevent
statistics) \cite{asipauskas}.
Here, grains have
variable size (grains grow with time, hence old grains in deep ice
are much larger than  young
grains near the dome surface): it would be difficult to select such a
fixed box size.

We thus chose a local averaging and define $\overline{\overline{M}}$
at each given
site as:
\begin{equation}
\overline{\overline{M}}
= \left\langle   \left(
\begin{array}{ll}
\ell_x^2 &\ell_x  \ell_z\\
\ell_z\ell_x & \ell_z^2
\end{array}
\right)\right\rangle_p
=   \frac{1}{N} \; \sum_{k=1}^N
\vec{\ell}(k)\otimes \vec{\ell}(k)
=
     {\cal R}
\left(
	\begin{array}{cc}
		 \lambda_1  & 0\\
			0 &  \lambda_2
	\end{array}
\right)
     {\cal R}^{-1}.
\label{defM}
\end{equation}
     Here, $ {\cal R}$ is the rotation which diagonalizes
$\overline{\overline{M}} $, and
$(\lambda_1,\lambda_2)$ are the corresponding eigenvalues;
$\left\langle \cdot
\right\rangle_p$ denotes the average over $N$ vectors, up to the
$p$-th neighbours (Fig. \ref{polariseurs}a). Hence, at each site
we include approximately the same number of vectors, $N\approx 3 +
6 + ... + 3\times2^{p-1}$. A smaller $p$  explores local details, a
larger $p$ (hence a larger scale)
improves the statistics.

\begin{figure}
\onefigure[ width=13cm]{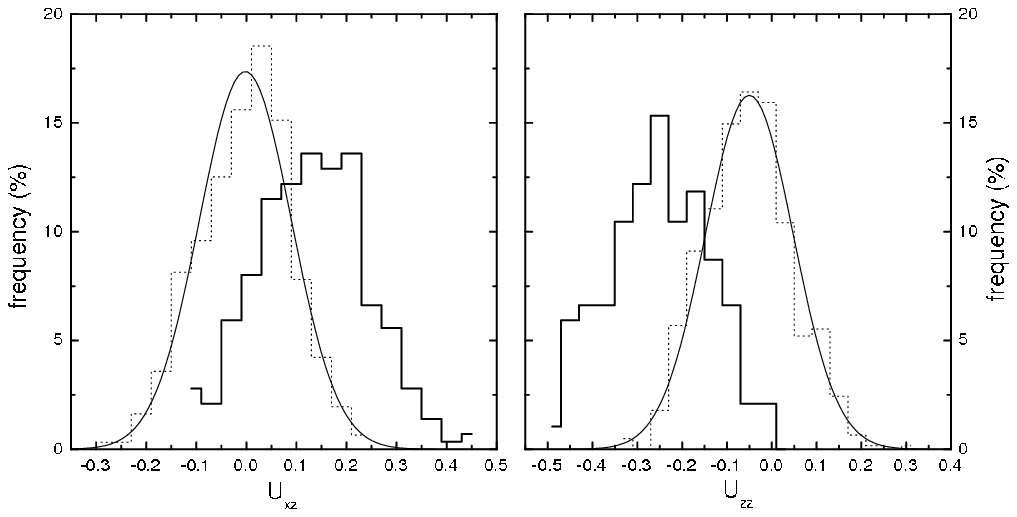}
\caption{Histogram of $U_{xz}$ (left) and of $U_{zz} (right)$
     measured with $p=3$ on Fig.
(\protect\ref{polariseurs}b):  362 m deep sample (dots), and the
corresponding gaussian distribution
(same mean and standard deviation) (thin solid line);
and on
Fig.
(\protect\ref{polariseurs}d):  2629  m deep sample (thick solid line).
} \label{Fig_angular}
\end{figure}

This tensor, independent from cristallographic information,
does not require any  knowledge or hypothesis regarding the past
history of the material. The diagonal components  $M_{xx}$ and
$M_{zz}$ of this tensor are both of order of the average square
distance $\left\langle \ell ^{2}\right\rangle $ between sites.
Conversely, the off-diagonal component $M_{xz}=M_{zx}$ is much
smaller, and even vanishes when the pattern
is isotropic. Hence  $\lambda_1 , \lambda_2 >0$: the
largest eigenvalue corresponds to the  direction in which grains are
most elongated
(Fig. \ref{polariseurs} b and d).

While many other
quantitative descriptors exist \cite{mecke}, few are adapted to
the determination of the anisotropy of such microstructure pattern.
Previous studies
on polar ice calculate the aspect ratio of the grains from the linear intercept
method \cite{gay,Arnaud}. Other use the eigenvalues of the grain inertia tensor
\cite{azuma,Wilen}; this is a true tensor, with mathematical
advantages: for instance,
its value does not depend on the particular choice of axes, hence
undergoes less artifacts when examining the 2D cut of a 3D pattern.

The tensor $\overline{\overline{M}}$ has the same advantages; but
also an additional one, thanks to its definition (eq.  \ref{defM})
being quadratic in $\vec{\ell}$: it has a physical signification in
terms of mechanical deformations \cite{aubouy}. More precisely, its
variations (with respect to a reference ${\overline {\overline
{M_{0}}}}$ measured in an isotropic, relaxed state) define a statistical strain
tensor
$\overline{\overline{U}}$:
\begin{equation}
		\overline{\overline{U}}
	= \frac{ \log \left(
       		{\overline {\overline {M}}}\right) -
		\log \left(
       		{\overline {\overline {M_{0}}}}
	 \right) }{2},
	\label{textdef}
\end{equation}
which exists even during large deformations and, at least  during
elastic deformations, coincides with the classical definition of
strain \cite{aubouy}.

For the 2D pattern studied here, $\overline{\overline{M_{0}}} $ is
     unknown, and processes other than deformation (such as grain
growth) modify the
microstructure. We first assume that the (viscoplastic) deformation
is isochore (incompressibility of ice),
so that $Tr \overline{\overline{U}}=0$ and $\lambda_1^0 \lambda_2^0 =
\lambda_1 \lambda_2$.
We also
assume that the reference state
     is isotropic;  hence  $\overline{\overline{M_{0}}} $ is isotropic
too, so that
$\lambda_1^0 =  \lambda_2^0 = \sqrt{\lambda_1\lambda_2}$.
We then can measure $\overline{\overline{U}}$ through the following equation:
          \begin{equation}
\overline{\overline{U}}
=
\frac{1}{2}   {\cal R}
\left(
	\begin{array}{cc}
		\log \left(\frac{\lambda_1}{\sqrt{\lambda_1
\lambda_2}}\right) & 0\\
			0 & \log
\left(\frac{\lambda_2}{\sqrt{\lambda_1\lambda_2}}\right)
	\end{array}
\right)
     {\cal R}^{-1}
=
\frac{1}{4}  {\cal R}
\left(
	\begin{array}{cc}
		\log \left(\frac{\lambda_1}{  \lambda_2}\right) & 0\\
			0 & \log \left(\frac{\lambda_2}{\lambda_1}\right)
	\end{array}
\right)
{\cal R}^{-1}.
\label{def}
\end{equation}
Hence $\overline{\overline{U}}$ has two independent components
$U_{zz}$ and $U_{xz}$, both largely insensitive to the grain size
fluctuations.

To estimate the statistical variability of the measure, we generate
$150$ isotropic
microstructures from a 2D Potts model of normal grain growth \cite{Potts}.
We check that  (i) each component of ${\overline {\overline U}}$
fluctuates around zero;
(ii)
its distribution is  gaussian (see an example with $p=3$ on Fig.
\ref{Fig_angular});
(iii)
     its standard deviation $\sigma$
only depends on the number $N$ of vectors, {\it i.e} on the scale of
observation
$p$;
and (iv)
     $\sigma(N) \propto N^{-1/2}$: more precisely, $\sigma \approx  0.34
\; N^{-1/2}$.

\section{Results}

\begin{figure}
\onefigure[width=14cm]{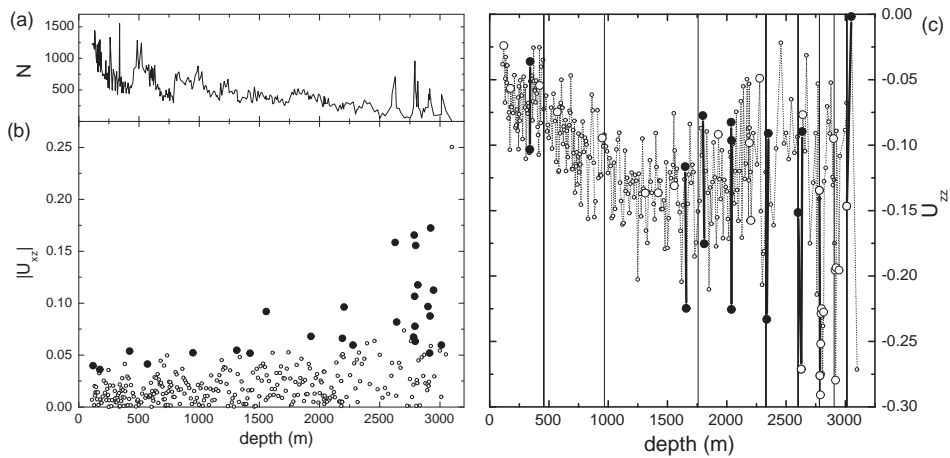}
\caption{
Localization of deformation
inside the ice core. For each point $\sigma$ is deduced from the
number of vectors $N$ (a).
Each point
corresponds to one of  329 samples; same scale for both figures.
(b)
$\vert U_{xz}\vert $ {\it versus} depth.
Closed circles highlight samples
where $\vert U_{xz}\vert  > 3 \sigma$.
(c)
$U_{zz}$ {\it versus} depth. Two large closed circles linked by a
thick solid line indicate
two successive measures
which 3 $\sigma$ confidence
intervals do not overlap.
A large open circle indicates a sample with a large significant shear
($U_{xz} >
3 \sigma$, from (b)).
Vertical lines indicate
zones, associated to climatic transitions, where the grain
growth rate changes significantly (see Discussion).
}
\label{Fig_depth}
\end{figure}

We first measure the deformation and its heterogeneity within one sample.
Fig. (\ref{polariseurs}b) shows that  the upper sample (depth 362 m)
is almost homogeneous. Its anisotropy
is small and, as  expected from a sample taken at
a dome, it mainly reflects a uniaxial compression along the vertical axis.
Conversely  (Fig. \ref{polariseurs}d), the anisotropy of the
lower sample, at 2629 m: is stronger; is localized in space;
and   breaks the axisymmetry around $z$ (see Discussion).
This symmetry breaking cannot result from the tilt of the ice core,
as the grains are tilted in average by 13.5$^\circ$ from the vertical,
much more than the core tilt at 2629 m (2.6$^\circ$).

Quantitatively,  the distributions
of $U_{xz}$ (vertical compression) and $U_{zz}$
(horizontal shear) of both samples differ significantly (Fig.
\ref{Fig_angular}).
The upper sample is
not sheared ($\langle U_{xz} \rangle = 0.00$),
and is
slightly flattened onto the horizontal plane ($\langle U_{zz} \rangle
= -0.05$).
The lower sample is
significantly sheared ($\langle U_{xz} \rangle = 0.13$)
and
strongly flattened ($\langle U_{zz} \rangle = -0.25$).
     Note that for $p=3$ ($N=21$) the
intrinsic value of $\sigma$ (see above) is  0.074; here, the
statistical deviation ($0.1$) is comparable,
but slightly larger, reflecting the  heterogeneity of deformation at
small scales
($10^{-2} -10^{-1}$ m).

We now turn to heterogeneities at large scale ($1 -10^{2}$ m), by
measuring  the global $\overline{\overline{U}} $ (integrated over all
the grain boundaries of a given sample), {\it versus} the sample
depth (Fig. \ref{Fig_depth}).
Whereas $U_{xz}$ remains
around zero, as expected in a dome situation, $U_{zz}$  increases
(in absolute value) with depth.

The essential point is the localization:
namely, the observation of a deformed (sheared or compressed) layer
immediately close to less deformed ones.
As mentioned, the grain size increases with
depth, therefore
     the number of vectors per sample
decreases \ref{Fig_depth}a, hence $\sigma$
increases. The localization is nevertheless significant, and  is
increasingly frequent at increasing depth (closed circles on Fig.
\ref{Fig_depth}b and c).

\section{Discussion}

Since the pictures are 2D cuts of actual 3D grains, the apparent
average grain size
could be misestimated; however, since $\overline{\overline{U}} $ is
dimensionless, it should remain unaffected.
When, and
only when, the pattern is axisymmetric, we can refine the above
analyzis (eq. \ref{def}). In that case,  the eigenvectors are
$(x,y,z)$, and the corresponding eigenvalues are
$(\lambda_1,\lambda_1,\lambda_2)$. The eigenvalues of
$\overline{\overline{U}}$ are now $\log(\lambda_1/\lambda_2)/6$
and $\log(\lambda_2/\lambda_1)/3$.
We have checked that the results corresponding to the 2D analyzis
(Figs. \ref{Fig_angular}, \ref{Fig_depth}) and the 3D one (data not
shown) are completely similar, with even slightly more heterogeneities
using the latter analysis.
Of course, care is required for the interpretation of an individual
sheared sample  (for instance regarding $U_{zz}$ at 2629 m, Fig.
\ref{Fig_angular}b). But on average,
the breaking of the axisymmetry hypothesis (large open circles on
Fig. (\ref{Fig_depth}c)) does not introduce significant artifacts on the
compression heterogeneities we recorded (closed circles on Fig.
\ref{Fig_depth}c).

A smooth shear deformation increasing with depth, sampled with a random
azimuth, could lead to an apparent heterogeneity of both $U_{zz}$ and $U_{xz}$.
However, this would imply an increasing anisotropy of the microstructure on
the horizontal $xy$ plane, whereas six horizontal thin sections were 
also analyzed (very few were available)
and showed very small anisotropy. This isotropy of the horizontal 
thin sections analyzed,
combined with the $U_{xz}$ values highlighted on figure 
\ref{Fig_depth}b is actually another
expression of strain heterogeneity.

Processes such as normal grain growth counteract the anisotropy of grains
induced by deformation.
Hence the strain recorded
by the microstructure $\overline{\overline{U}}$  underestimates the
strain $\overline{\overline{\varepsilon}}$ actually experienced
by the material.
The growth rate itself could undergo some fluctuactions.
At few depths, near the bottom of the core, a correlation has been observed
between small grain sizes (related to a large amount of dust in the
ice \cite{Weiss})
and large strains. In these few cases, highlighted by the thick
vertical lines in figure \ref{Fig_depth}c,
the large differences of $U_{zz}$ observed between two adjacent
layers could be partly explained by the difference in
the grain growth rate.
    However, this effect is unable to explain most
heterogeneities (closed circles on Fig. \ref{Fig_depth}c). In
addition, it does not explain the large number of shear layers
(closed circles on Fig. \ref{Fig_depth}b).
We note here that such a correlation between grain size and $U_{zz}$ 
would be difficult
to explain in case of a strong effect of a random azimuth.
We thus argue that the heterogeneity of anisotropy, observed at
both small and large scales, cannot be entirely an observation 
artifact and has a mechanical origin.

\section{Conclusion and perspectives}

A dating
chart is the relation $t(z)$ between depth
$z$ and age $t$ of ice:
it requires a model, and hypotheses.
The ice core is drilled exactly at the summit of a dome,
in order to assume  that the flow is axisymmetric; so that the
ice thinning results from vertical compression only, without horizontal shear.
Current models further assume
a smooth and
monotonous increase of the thinning  of ice defined as
the ratio
$e(z) / e(0)$,
where $e$ is the thickness of the annual ice layer
at depth $z$.

The second assumption has already been shown to be wrong in the
Greenland ice cores GRIP and GISP2
(\cite{alley,Gow,Dorthe}).
Flow disturbances have been reported
within at least the $30\%$ deepest part of the GISP2 core, based on
the observations
of wavy ash layers,   crystal stripings visible by eyes,
and anomalous fabrics \cite{alley}.
The dating has been particularly questioned by the observation of a
folding, {\it i.e.} local inversion of
ice layers  \cite{Grootes}.

Here, our method is more accurate and applies to ice itself, without
requiring markers nor extreme events.
We
show that flow disturbances are detectable from almost the top of the
ice sheet,
     and increase in number and intensity with depth. In fact, both dating model
assumptions are contradicted by our results: the strain gradient  is
variable and not even always positive
(Fig. \ref{Fig_depth}c); and although the flow at a dome is
axisymmetric on average, in detail there is a symmetry breaking due
to shear, especially in deep layers (Fig. \ref{Fig_depth}b).
This suggests to reconsider current standard dating
charts.


In the future, we expect to correlate the grain boundary pattern with 
the $c$-axis orientation, to improve
our understanding of their coupling.

      {  \acknowledgments We thank M. Aubouy for  his support and useful
discussions, M. Glock for reading the manuscript, and M. Krinner and
M. Manouvrier for
their technical support.
        This work was partially supported by CNRS ATIP 0693.
This work is a contribution to the ``European project for Ice
Coring in Antarctica" (EPICA), a joint ESF (European Science
Foundation)/EC scientific program, funded by the European
Commission under the Environment and Climate Programme (1994-1998)
contract ANV4-CT95-0074 and by national contributions from
Belgium, Denmark, France, Germany, Italy, the Netherlands, Norway,
Sweden, Switzerland and the United Kingdom.}

\end{document}